\documentclass[12pt,preprint]{emulateapj}
\usepackage{graphicx}
\usepackage{xcolor}

\shorttitle{New Analysis of HD~6434}
\shortauthors{Hinkel et al.}
\slugcomment{Submitted for publication in the Astrophysical Journal}

\def\gtaprx{ \mathrel{ \vcenter{
      \offinterlineskip \hbox{$>$}
      \kern 0.3ex \hbox{$\sim$}    } } }

\def\ltaprx{ \mathrel{ \vcenter{
      \offinterlineskip \hbox{$<$}
      \kern 0.3ex \hbox{$\sim$}    } } }
      
\def\aj{{AJ}}
\def\apj{{ApJ}}
\def\apjs{{ApJS}}

\def\aap{{A\&A}}
\def\pasp{{PASP}}

\def\mnras{{MNRAS}}

\begin{document}

\title{A New Analysis of the Exoplanet Hosting System HD~6434}

\author{
  Natalie R. Hinkel\altaffilmark{1,2},
  Stephen R. Kane\altaffilmark{1},
  Genady Pilyavsky\altaffilmark{2},
  Tabetha S. Boyajian\altaffilmark{3},
  David J. James\altaffilmark{4},
  Dominique Naef\altaffilmark{5},
  Debra A. Fischer\altaffilmark{4},
  Stephane Udry\altaffilmark{5}
}
\email{natalie.hinkel@gmail.com}
\altaffiltext{1}{Department of Physics \& Astronomy, San Francisco
  State University, 1600 Holloway Avenue, San Francisco, CA 94132,
  USA}
\altaffiltext{2}{School of Earth \& Space Exploration, Arizona State University,
  Tempe, AZ 85287, USA}
\altaffiltext{3}{Department of Astronomy, Yale University, New Haven,
  CT 06511, USA}
  \altaffiltext{4}{Cerro Tololo Inter-American Observatory, Colina el Pino, La Serena, Chile}
\altaffiltext{5}{Observatoire de Genve, 51 Ch. des Maillettes, 1290 Sauverny, Switzerland}

\begin{abstract}
The current goal of exoplanetary science is not only focused on detecting but characterizing planetary systems in hopes of understanding
how they formed, evolved, and relate to the Solar System. The Transit Ephemeris Refinement and Monitoring Survey (TERMS) combines both radial velocity (RV) and photometric data in order to achieve unprecedented ground-based precision in the fundamental properties of nearby, bright, exoplanet-hosting systems. Here we discuss HD~6434 and its planet, HD~6434b, which has a $M_p$\,sin\,$i$ = 0.44 $M_J$ mass and orbits every 22.0170 days with an eccentricity of 0.146. We have combined previously published RV data with new measurements to derive a predicted transit duration of $\sim$6 hrs, or 0.25 days, and a transit probability of 4\%. Additionally, we have photometrically observed the planetary system using both the 0.9m and 1.0m telescopes at the Cerro Tololo Inter-American Observatory, covering 75.4\% of the predicted transit window. We reduced the data using the automated TERMS Photometry Pipeline, developed to ensure consistent and accurate results. We determine a dispositive null result for the transit of HD~6434b, excluding the full transit to a depth of 0.9\% and grazing transit due to impact parameter limitations to a depth of 1.6\%.  

\end{abstract}


\section{Introduction}
\label{introduction}

The HD~6434 system has been of interest to observers for $\sim$15 years. In \citet{Santos01}, the abundances of the host-star were analyzed in detail in order to better understand the planet-iron correlation. Surprisingly they found that, despite having a confirmed exoplanet, the star was not only deficient with respect to the photospheric iron abundance, [Fe/H], but the most deficient ([Fe/H] = $-$0.55 $\pm$ 0.07) out of the 21 stars with planets and brown dwarf companions in their sample. 

The planet orbiting HD~6434 was described in great detail in \citet{Mayor04}, as a member of the {\sc CORALIE} planet-search program in the southern hemisphere \citep{Udry00}.  The radial velocity (RV) measurements allowed \citet{Mayor04} to calculate the most precise Keplerian orbital solution at the time for the planet. They noted that, compared to the other 12 single-planet systems in their dataset, HD~6434b had not only a shorter period but also a lower eccentricity and relatively low mass, namely $M_p$\,sin\,$i$ = 0.397 $M_{J}$, or Jupiter masses. The HD~6434 system was also observed by \citet{MoroMartin07} using {\it Spitzer}'s Multiband Imaging Photometer, Infrared Spectrograph, and Infrared Array Camera in hopes of finding a dusty, debris disk around the known exoplanet host. However, their study did not reveal any correlation between debris disks and systems with hot-Jupiter-type planets, or with enhanced stellar [Fe/H] content.

We have re-observed the HD~6434 planetary system as part of the Transit Ephemeris Refinement and Monitoring Survey \citep{Kane09}, or TERMS, which observes stars with known RV planets in an effort to detect new transiting exoplanets and refine existing orbital parameters.  Through our international collaboration, we are able to utilize new RV and photometric data in an effort to improve both the stellar and planetary characteristics \citep{Kane15, Hinkel15, Wang12}. Our highly refined orbital ephemerides allows us to calculate a predicted transit window on the order of hours for targets with periods greater than 10 days, a feat not currently duplicated by any ground- or space-based missions \citep{vonBraun09}. While both the Kepler/K2 mission and Transiting Exoplanet Survey Satellite (TESS), set to launch in 2017, are primed to discover many transiting planets in the 10-90 day period regime, the TERMS project focuses on nearby, bright hosts with longer-period and/or highly eccentric planetary orbits. 

In this paper, we will present updated stellar properties as observed by the CHIRON instrument as well as stellar abundance information via the {\it Hypatia Catalog} \citep{Hinkel14} in Section \ref{stellar}. An extensive RV data set featuring both literature values and new data is discussed in Section \ref{keplerian}, along with updated Keplerian orbital solutions. We calculated the host star HD~6434 properties, and predict a transit depth of 0.9\% and a 1$\sigma$ transit window of 0.69 days. In the Appendix, we introduce the TERMS Photometry Pipeline (TPP), a photometric data reduction module that was created by our team to aid in the efficiency of reducing and analyzing TERMS data. The TPP was utilized for the CTIO 1.0m and 0.9m observations of HD~6434 that we have procured, discussed in Section \ref{photometry}, which ultimately revealed dispositive null result for the full transit exclusions of HD~6434 .

\section{Stellar Properties}
\label{stellar}
Using a variety of telescopes and data, we determine the properties of the host star, HD~6434, such that we may better understand the planetary companion.

\subsection{Fundamental Parameters}
\label{sec:sme}

The methodology of the TERMS project requires that we have an adequate understanding the host star properties, particularly the mass and radius.  To characterize HD~6434, we obtained a high-resolution spectrum using the CHIRON high resolution spectrograph installed on the Cerro Tololo Inter-American Observatory (CTIO) 1.5m telescope on 2014 July 7 \citep{Tokovinin13,Brewer14}. A 400~s exposure, between 4200--8800 \AA\,, was taken of the object with a resolution of $R = 90,000$, resulting in a signal to noise of $\sim 100$.

We use the Spectroscopy Made Easy (SME) package \citep{Valenti:2005p1491} in iterative mode \citep{Valenti09} to determine the stellar properties.  The iterative approach determines stellar properties from the observed spectrum and compares them to those predicted by the Yonsei-Yale model evolutionary isochrones \citep{Demarque04} in order to arrive at an internally consistent solution. The stellar values for HD~6434 are shown in Table \ref{starparams}, including both referenced and derived parameters from our CHIRON spectra analysis. The stellar temperature derived from spectral synthesis and the radius derived from evolutionary modeling are within 1-$\sigma$ agreement with those estimated from empirically calibrated ($V-J,H,K$) color - temperature relations, $T_{\rm eff}$ = 5739~K, and surface brightness relations, $R_{*, SB} = 1.10 \pm 0.04$~R$_{\odot}$ \citep{Boyajian2013, Boyajian14}. The derived parameters are consistent with the star being a G1IV m-2 star per \citet{Gray89}, who specifically targeted metal poor stars. 

\begin{deluxetable}{lcc}
  \tablecaption{\label{starparams} Stellar Parameters}
  \tablehead{
    \colhead{Parameter} &
    \colhead{Value} &
    \colhead{Source}
  }
 \startdata
  $V$      & $7.71 \pm 0.12$ & \citet{Carney78} \\
  $B-V$      & $0.61 \pm 0.01$ & \citet{Carney78}  \\
  Distance (pc)      & $41.4 \pm 1.03$   & Hipparcos \\
   Age (Gyr) & $11.27 \pm 2.12$ & SME \\
   $[$Fe/H$]$ (dex) & $-$0.48 $\pm$ 0.05 (\it{0.35}) & Hypatia \\
  $T_{\rm eff}$ (K)    & $5690 \pm  92$    & SME  \\
  $\log g$ (cgs)            & $4.29 \pm 0.06$   & SME     \\
  $v \sin i$ (km\,s$^{-1}$)    & $2.2  \pm 0.5$  & SME  \\
  $M_*$ (M$_{\odot}$)            & $0.89 \pm 0.04$ & SME  \\
  $R_*$ (R$_{\odot}$)            & $1.14 \pm 0.05$ & SME
  \enddata
\end{deluxetable}

In comparison to \citet{Santos01}, we find that our SME $T_{\rm eff}$ is 100 K lower and $\log g$ is 0.27 lower. Similar to \citet{Santos01}, \citet{Mayor04} had a larger $T_{\rm eff}$ value of 5835 K, $\log g$ of 4.60, similar $v \sin i$ of 2.3 km\,s$^{-1}$, and a smaller $M_*$ estimate of 0.79 M$_{\odot}$. \citet{MoroMartin07} also had a higher value for $T_{\rm eff}$, the same as \citet{Mayor04}, although their stellar mass estimate was between ours and \citet{Mayor04} at 0.84 M$_{\odot}$.

\subsection{Stellar Abundances}
\label{sec:abund}

To date, more than a dozen different groups, such as \citet{Ecuvillon:2004p2198} and \citet{Battistini15}, have measured +23 elemental abundances within the photosphere of the HD~6434 host star. Every group that measures stellar abundances normalizes their data with respect to a solar scale, however, those scales vary and therefore makes comparing multiple data sets difficult. Therefore, we renormalized the solar abundance scales for all of the datasets to be on a consistent scale with respect to \citet{Lodders:2009p3091}, per the analysis within the {\it Hypatia Catalog} \citep{Hinkel14}. The median [Fe/H] determination in HD~6434 was $-$0.48 dex, with a variation of 0.35 dex between the groups. The maximum, renormalized [Fe/H] measurement originated from \citet{Gilli:2006p2191} with $-$0.31 dex while the minimum measurement was from \citet{Laird:1985p1923} with $-$0.66 dex. Comparatively, the SME [Fe/H] determination for HD 6434 performed here is $-$0.61 $\pm$ 0.07 dex when renormalized. The standard error on [Fe/H] is $\sim$ 0.05 dex, meaning that the {\it spread} between groups is a factor of over five times larger than the typical error. The large group-to-group discrepancy within the literature, due to the variations between the methods and systematic offsets, means that HD~6434 was not included in the analysis or reduced version of the Hypatia Catalog \citep{Hinkel14}.

Many elements that are considered important for habitability, such as carbon, oxygen, magnesium, and silicon, were measured within the photosphere of HD~6434, namely [C/Fe] = 0.22 $\pm$ 0.09 ({\it 0.43}) dex, [O/Fe] = 0.35 $\pm$ 0.09 ({\it 0.28}) dex, [Mg/Fe] = 0.3 $\pm$ 0.07 ({\it 0.03}) dex, and [Si/Fe] = 0.13 $\pm$ 0.05 ({\it 0.07}) dex, where both the typical errors are included along with the respective spreads in parenthesis. In addition to [Mg/Fe], there were an additional ten elements for which the spread was smaller than the average error for that element: [N/Fe] = 0.08 dex, [S/Fe] = $-$0.14 dex, [ScII/Fe] = 0.025, [Cu/Fe] = $-$0.06 dex, [Zn/Fe] = 0.0 dex, [Y/Fe] = 0.24 dex, [YII/Fe] = 0.27 dex, [Zr/Fe] = 0.27 dex, [ZrII/Fe] = 0.32 dex, [BaII/Fe] = 0.16 dex, [Nd/Fe] = 0.13 dex, and [Eu/Fe] = 0.15 dex. In general, the volatile and neutron-capture element-ratios within HD~6434 were super-solar. However, element-ratios for those at the iron-peak were markedly sub-solar. While this conclusion may prove interesting when considering the presence of the b-planet around the host star, we must remember that the spread in the individual elements, in conjunction with the large spread in iron, makes any firm interpretation tentative until the methodologies can be better understand.

\subsection{Stellar Variability}
To examine the stellar variability of HD~6434, we performed a weighted Lomb-Scargle (L-S) Fourier analysis of our photometry (see \citet{Kane07} for more details). The L-S Fourier analysis was first applied to the Hipparcos photometry by \citet{Mayor04}, and like them, we do not find any peaks of significance at the measured orbital period of the planet. There is a peak of moderate significance at 24.13~days, however  given the evolved G1 classification of the star (see Section \ref{sec:sme}), it could possibly be due to the stellar rotation period. 

\begin{figure}
  \includegraphics[width=9.6cm]{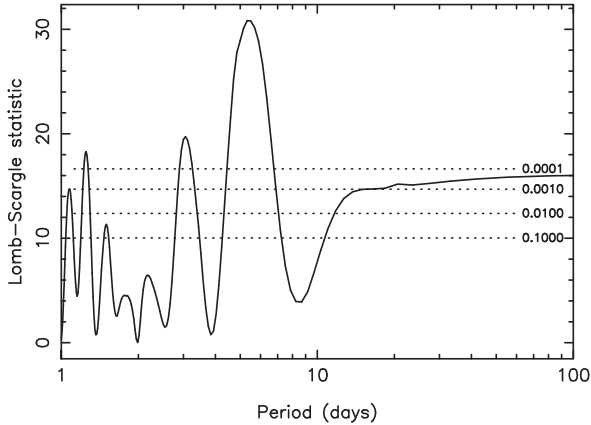}
  \caption{
  A weighted L-S periodogram of the HD~6434 CTIO 1.0m photometry. A periodic signature close to 5 days persists, as also seen in the fourier analysis of the Hipparcos photometry. The horizontal lines indicate significance values starting at 10\% false-alarm probability and going up to 0.01\% FAP.
  }
  \label{pgram}
\end{figure}

The strongest peak in the Fourier analysis of the Hipparcos photometry is at 5.12~days. A Fourier analysis of our CTIO 1.0m photometry (see Figure \ref{pgram}) are unable to confirm the 24~day period since the data sampling do not cover a complete orbital period of the planet. The power spectrum does, however, reveal a significant periodic signature at 5.32~days, similar to the Hipparcos signature at 5.12~days. The nature of such a periodic signature is difficult to explain from a stellar astrophysics perspective and requires further study. 
The horizontal lines in Fig. \ref{pgram} indicate significance values starting at false-alarm probability (FAP) ranging from 10-0.01\% FAP. The FAP are extremely low for these calculations because the number of data points is high and the range of frequencies that we sample in the Fourier analysis is large. In addition, the FAP is slightly underestimated due to non-Gaussian sources of noise in the photometry, although this effect is minimal.
As a final check, we also examined a single night of CTIO 0.9m photometry for less than 0.5 day periodic signatures but no features of significance were detected in the Fourier analysis.

\section{Keplerian Orbit and Transit Ephemeris}
\label{keplerian}

Previous RV observations of HD~6434 \citep{Mayor04} were undertaken
with the {\sc CORALIE} spectrometer mounted on the 1.2-m Euler Swiss
telescope at La Silla. Those observations were conducted as part of the
{\sc CORALIE} exoplanet search program \citep{Udry00}. We continued to
monitor HD~6434 using {\sc CORALIE} to improve the Keplerian orbital
solution and provide an accurate transit ephemeris. The complete
dataset of 137 measurements is shown in Table \ref{rvs} including
previously acquired measurements and 59 new measurements. These new
measurements extend the overall time baseline for the RV observations by
a factor of $\sim$3.6, for a total baseline of $\sim$15 years. Due 
to a major instrument upgrade that occurred in 2007, the data are
divided into two separate datasets indicated by
the horizontal lines in Table \ref{rvs}. As a
result, the older data cannot be reduced with the most recent
reduction pipeline version because of the lower quality of the
calibrations. After the 2007 upgrade, the quality requirements for the
calibrations were increased as well as the science data signal-to-noise (S/N). It is
also worth noting that the stated RV errors include an instrumental
systematic error term that has been added in quadrature to the photon
noise error. These terms are 0.006 and 0.005 km\,s$^{-1}$ for the old
and new {\sc CORALIE} datasets respectively.

\LongTables
\begin{deluxetable}{ccc}
  \tablewidth{0pc}
  \tablecaption{\label{rvs} HD~6434 Radial Velocities}
  \tablehead{
    \colhead{Date} &
    \colhead{RV} &
    \colhead{$\sigma$} \\
    \colhead{(BJD -- 2,440,000)} &
    \colhead{(km\,s$^{-1}$)} &
    \colhead{(km\,s$^{-1}$)}
  }
  \startdata
 11142.68020 & 22.9974 & 0.0089 \\
 11432.80027 & 23.0331 & 0.0111 \\
 11446.75913 & 22.9899 & 0.0096 \\
 11454.73999 & 23.0329 & 0.0082 \\
 11464.69787 & 23.0199 & 0.0082 \\
 11480.67298 & 23.0421 & 0.0092 \\
 11485.65180 & 23.0261 & 0.0076 \\
 11490.59534 & 22.9926 & 0.0089 \\
 11495.57333 & 23.0074 & 0.0074 \\
 11496.57667 & 23.0148 & 0.0082 \\
 11497.62586 & 23.0299 & 0.0078 \\
 11498.63156 & 23.0470 & 0.0073 \\
 11503.57269 & 23.0447 & 0.0075 \\
 11541.57731 & 23.0405 & 0.0092 \\
 11550.57201 & 23.0513 & 0.0101 \\
 11551.59933 & 23.0399 & 0.0087 \\
 11552.55969 & 23.0340 & 0.0080 \\
 11554.56092 & 22.9973 & 0.0082 \\
 11562.54747 & 23.0215 & 0.0084 \\
 11568.55132 & 23.0659 & 0.0092 \\
 11570.54645 & 23.0498 & 0.0077 \\
 11571.56369 & 23.0513 & 0.0095 \\
 11573.53642 & 23.0365 & 0.0095 \\
 11576.53597 & 23.0269 & 0.0105 \\
 11578.54635 & 22.9982 & 0.0109 \\
 11579.54676 & 22.9681 & 0.0081 \\
 11581.57223 & 22.9862 & 0.0087 \\
 11583.53548 & 22.9923 & 0.0076 \\
 11585.52716 & 23.0011 & 0.0087 \\
 11587.52952 & 23.0316 & 0.0083 \\
 11619.50284 & 23.0109 & 0.0108 \\
 11748.88635 & 23.0547 & 0.0093 \\
 11749.89176 & 23.0360 & 0.0074 \\
 11750.89718 & 23.0207 & 0.0076 \\
 11751.91750 & 23.0120 & 0.0075 \\
 11752.92224 & 23.0022 & 0.0077 \\
 11753.89216 & 22.9954 & 0.0074 \\
 11754.91855 & 22.9897 & 0.0086 \\
 11755.89178 & 22.9821 & 0.0082 \\
 11756.88234 & 22.9766 & 0.0082 \\
 11757.92253 & 22.9973 & 0.0082 \\
 11758.88486 & 22.9987 & 0.0077 \\
 11759.91148 & 23.0064 & 0.0079 \\
 11774.86156 & 22.9932 & 0.0076 \\
 11776.80808 & 22.9853 & 0.0081 \\
 11777.85738 & 22.9814 & 0.0082 \\
 11778.82124 & 22.9844 & 0.0089 \\
 11780.70609 & 22.9899 & 0.0089 \\
 11782.85670 & 23.0128 & 0.0078 \\
 11783.72682 & 23.0137 & 0.0078 \\
 11784.71047 & 23.0340 & 0.0077 \\
 11785.78082 & 23.0416 & 0.0076 \\
 11786.81941 & 23.0511 & 0.0075 \\
 11788.82536 & 23.0500 & 0.0110 \\
 11789.74667 & 23.0609 & 0.0073 \\
 11790.83167 & 23.0405 & 0.0076 \\
 11791.66142 & 23.0704 & 0.0079 \\
 11793.73954 & 23.0461 & 0.0090 \\
 11804.81570 & 23.0232 & 0.0077 \\
 11805.78097 & 23.0219 & 0.0095 \\
 11806.76512 & 23.0415 & 0.0087 \\
 11836.70969 & 23.0430 & 0.0075 \\
 11837.72282 & 23.0312 & 0.0076 \\
 11838.69306 & 23.0069 & 0.0078 \\
 11861.61053 & 23.0219 & 0.0076 \\
 11868.63823 & 23.0081 & 0.0075 \\
 11899.54644 & 23.0548 & 0.0087 \\
 11901.60605 & 23.0532 & 0.0079 \\
 11912.54895 & 23.0046 & 0.0079 \\
 11915.56590 & 23.0201 & 0.0081 \\
 11918.56685 & 23.0537 & 0.0098 \\
 11922.55565 & 23.0624 & 0.0104 \\
 12129.88404 & 22.9842 & 0.0073 \\
 12162.77443 & 23.0531 & 0.0077 \\
 12190.77286 & 23.0247 & 0.0075 \\
 12221.64737 & 23.0142 & 0.0073 \\
 12486.84524 & 23.0133 & 0.0078 \\
 12643.54878 & 23.0406 & 0.0077 \\
 12845.91134 & 23.0448 & 0.0105 \\
 12879.86766 & 22.9907 & 0.0088 \\
 12886.73778 & 23.0427 & 0.0082 \\
 14056.56344 & 23.0520 & 0.0094 \\
 14063.52899 & 23.0316 & 0.0101 \\
 14066.55745 & 22.9889 & 0.0091 \\
 14071.54499 & 23.0228 & 0.0085 \\
 --- & --- & --- \\
 14467.61208 & 23.0147 & 0.0063 \\
 14710.84386 & 23.0193 & 0.0067 \\
 14734.78571 & 23.0540 & 0.0065 \\
 14810.57732 & 23.0383 & 0.0067 \\
 14814.61477 & 23.0098 & 0.0061 \\
 15106.77851 & 23.0207 & 0.0060 \\
 15796.84800 & 23.0753 & 0.0062 \\
 15840.71899 & 23.0553 & 0.0061 \\
 15857.62039 & 23.0554 & 0.0062 \\
 15859.56419 & 23.0597 & 0.0059 \\
 15862.66864 & 23.0777 & 0.0060 \\
 15864.71915 & 23.0639 & 0.0059 \\
 15866.68793 & 23.0634 & 0.0059 \\
 15868.58252 & 23.0383 & 0.0060 \\
 15870.56294 & 23.0098 & 0.0059 \\
 15872.57603 & 23.0094 & 0.0060 \\
 15874.61053 & 23.0119 & 0.0064 \\
 16111.85047 & 23.0323 & 0.0062 \\
 16112.88548 & 23.0187 & 0.0062 \\
 16113.85252 & 23.0025 & 0.0062 \\
 16114.85473 & 22.9990 & 0.0061 \\
 16115.86574 & 23.0106 & 0.0061 \\
 16116.87374 & 23.0080 & 0.0060 \\
 16117.86466 & 23.0175 & 0.0060 \\
 16118.86015 & 23.0232 & 0.0060 \\
 16134.78760 & 23.0292 & 0.0066 \\
 16141.88284 & 23.0391 & 0.0076 \\
 16172.79721 & 23.0659 & 0.0064 \\
 16195.73919 & 23.0743 & 0.0061 \\
 16229.64827 & 23.0421 & 0.0062 \\
 16231.67904 & 23.0606 & 0.0062 \\
 16235.70821 & 23.0670 & 0.0060 \\
 16236.61581 & 23.0674 & 0.0062 \\
 16460.94139 & 23.0645 & 0.0060 \\
 16471.90467 & 23.0341 & 0.0072 \\
 16472.87607 & 23.0506 & 0.0065 \\
 16474.94128 & 23.0572 & 0.0063 \\
 16475.94226 & 23.0623 & 0.0064 \\
 16477.91262 & 23.0798 & 0.0061 \\
 16481.83515 & 23.0684 & 0.0070 \\
 16486.94303 & 23.0192 & 0.0067 \\
 16530.90905 & 23.0292 & 0.0073 \\
 16545.69771 & 23.0746 & 0.0060 \\
 16553.61502 & 23.0138 & 0.0089 \\
 16577.67415 & 22.9972 & 0.0060 \\
 16578.61259 & 22.9973 & 0.0058 \\
 16579.67094 & 23.0176 & 0.0060 \\
 16581.77300 & 23.0378 & 0.0060 \\
 16583.72912 & 23.0470 & 0.0059 \\
 16584.75011 & 23.0569 & 0.0059 \\
 16585.62185 & 23.0674 & 0.0061 \\
 16586.74797 & 23.0737 & 0.0061 \\
  \enddata
\end{deluxetable}

\begin{figure*}
  \begin{center}
    \includegraphics[width=16.0cm]{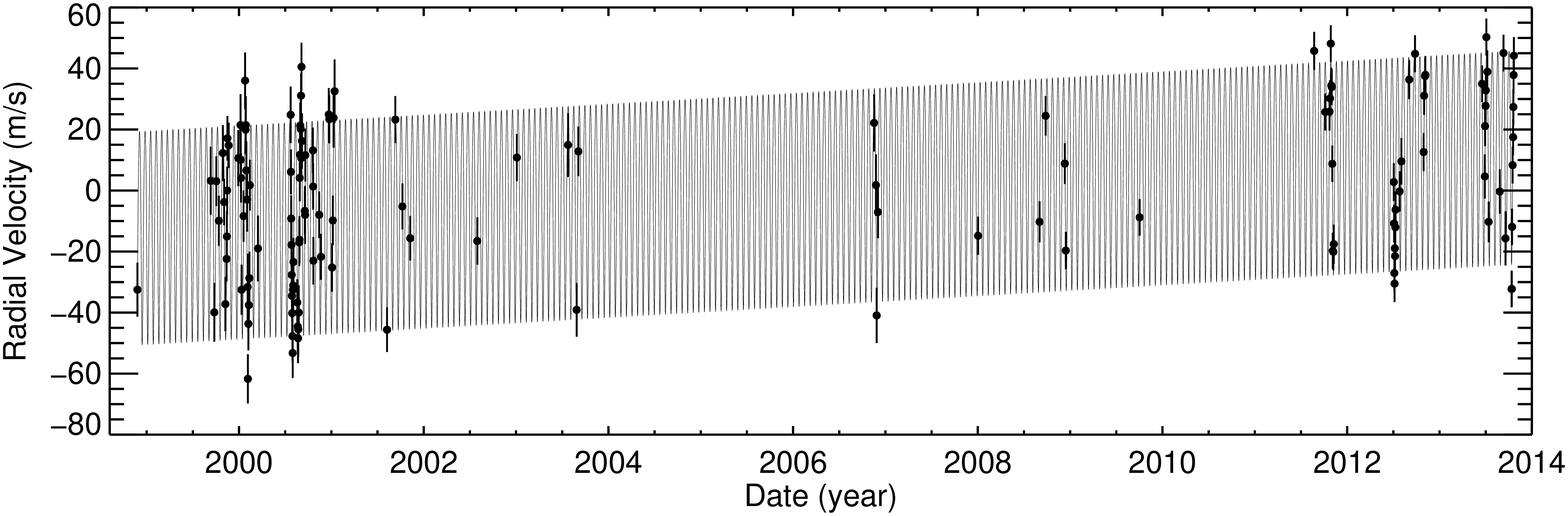} \\
    \includegraphics[width=16.0cm]{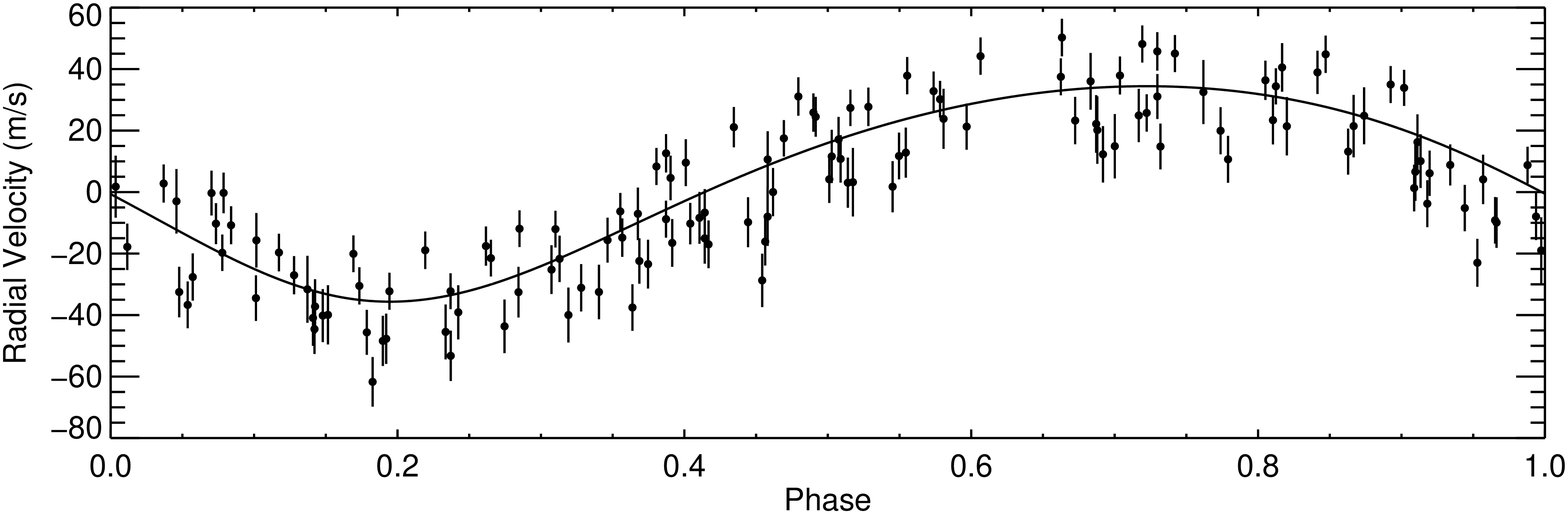}
  \end{center}
  \caption{{\it Top}: The complete RV time-series of HD~6434 including
    137 measurements acquired over 15 years (see Table \ref{rvs}). The
    solid line indicates the Keplerian orbital solution, shown in
    Table \ref{planet}. Note the inclusion of a linear RV trend which
    is now apparent with the improved time coverage. {\it Bottom}: The
    RV data and best-fit model phased on the orbital period of the
    planet.}
    \vspace{5mm}
  \label{rv}
\end{figure*}

The Keplerian orbital solution fit to the RV data in Table \ref{rvs} was
performed used {\sc RVLIN}: a partially linearized, least-squares fitting
procedure discussed in \citet{Wright09}. The uncertainties in the fit
parameters were estimated using the {\sc BOOTTRAN} bootstrapping routines
described in \citet{Wang12}. The orbital solution is shown in Table
\ref{planet}, with unphased and phased representations shown in Figure
\ref{rv}. The data in Figure \ref{rv} are folded on the best fit orbital period determined from the RVs, where the individual parameter errors and cumulative errors are taken into consideration via the \citet{Wang12} methodology.
The offset between the ``old'' (pre-upgrade) and ``new'' (post-2007)
{\sc CORALIE} data was included as a free parameter in the
fit, such that the code is able to produce the offset between the datasets, along with the Keplerian solution.
The new data have an RV offset of $-0.35 \pm 4.88$~m\,s$^{-1}$
relative to the old data, where
the RV offset is not included in Table \ref{rvs}. Typical stability of RV standard stars is 100-300 m/s, per \citet{Soubiran13}.
We also find strong evidence of a linear
trend in the RV data (see parameter $dv/dt$ in Table \ref{planet} and
the upper panel of Figure \ref{rv}), indicating a potential long-period companion in the system.
The fit without the linear trend has a $\chi^2_{\mathrm{red}}$ of 1.15 and an rms of 8.08. Given the relatively large uncertainties of the individual RV measurements, more RV data is needed to verify the strength of the RV trend and to verify that a ``turn-around" occurs.
Further RV monitoring of the HD~6434 system will also help to resolve the nature of the linear trend.
Shorter-term periodicities are likely aliases due to the observing cadence and two-planet RV-fits did not 
reveal any signatures with a realistic RMS. 

Our orbital solution matches both the period ($P$) and semi-major axis ($a$) in \citet{Santos01}, however, we a discrepancy between $M_p$\,sin\,$i$, where they estimated 0.48 $M_J$, and the eccentricity, where they determined $e =$ 0.295. The orbital parameters calculated in \citet{Mayor04} more closely resemble our values to within error. The planetary properties used in \citet{MoroMartin07} were taken from \citet{Mayor04}. We have significantly more data and time baseline compared to \citet{Mayor04} which greatly improves the constraints on the orbital shape and orientation. 

\begin{deluxetable}{lc}
  \tablecaption{\label{planet} Keplerian Orbital Model}
  \tablewidth{0pt}
  \tablehead{
    \colhead{Parameter} &
    \colhead{Value}
  }
  \startdata
\noalign{\vskip -3pt}
\sidehead{HD~6434 b}
~~~~$P$ (days)                    & $22.0170 \pm 0.0008$ \\
~~~~$T_c\,^{a}$ (BJD -- 2,440,000) & $16859.616 \pm 0.220$ \\
~~~~$T_p\,^{b}$ (BJD -- 2,440,000) & $11909.308 \pm 0.638$ \\
~~~~$e$                           & $0.146 \pm 0.025$ \\
~~~~$\omega$ (deg)                & $163.2 \pm 10.5$ \\
~~~~$K$ (m\,s$^{-1}$)             & $35.0 \pm 0.9$ \\
~~~~$M_p$\,sin\,$i$ (M$_J$)       & $0.44 \pm 0.01$ \\
~~~~$a$ (AU)                      & $0.148 \pm 0.002$ \\
~~~~$dv/dt$ (m/s/day)             & $0.005 \pm 0.001$\\
\sidehead{System Properties}
~~~~$\gamma$ (m\,s$^{-1}$)           & $3.07 \pm 2.59$ \\
\sidehead{Measurements and Model}
~~~~$N_{\mathrm{obs}}$            & 137 \\
~~~~rms (m\,s$^{-1}$)             & 7.77 \\
~~~~$\chi^2_{\mathrm{red}}$       & 1.01
  \enddata
  \tablenotetext{a}{Time of mid-transit.}
  \tablenotetext{b}{Time of periastron passage.}
\end{deluxetable}

The Keplerian orbital solution shown in Table \ref{planet} includes a
predicted time of inferior conjunction ($T_c$) which, if the orbital
inclination is close to $i = 90\degr$, corresponds to the time of
mid-transit. Future times of possible transit are calculated using a
Monte-Carlo bootstrap that propagates the uncertainty in this orbital
parameter to the time of the transit \citep{Wang12}. The combination of
the stellar properties in Table \ref{stellar} and the planetary
parameters in Table \ref{planet} may be used to calculate the
predicted properties of a potential transit. Using the mass-radius
relationship described by \citet{Kane:2012p8338}, we approximate the radius of
the planet as $R_p = 1.0$ R$_J$. These values result in a transit
probability of 4\%, a transit duration of 0.25~days, and a transit
depth of 0.9\%, including the effects of eccentricity
\citep{Kane08}. The 1$\sigma$ transit window for the planet is thus
0.69~days, or 3\% of the orbital period \citep{Kane09}.

\section{Sources of Photometry}
\label{photometry}

Using both the stellar and planetary properties given in Tables \ref{stellar} and \ref{planet}, respectively, we determined the window, duration, depth, and probability of a potential transit for the HD~6434's b-planet (see Section \ref{keplerian}). Because the 1$\sigma$ transit window is 0.69~days, we only required one night of photometric data in order to conclusively confirm or rule out the transit of the b-planet. However, we also took advantage of the publicly available Hipparcos data as well as out-of-window, baseline photometric data from CTIO.

\subsection{Photometry from Hipparcos}
\label{hipparcos}
We utilized publicly available data from the {\it Hipparcos} satellite. {\it Hipparcos} acquired a
total of 142 measurements of HD~6434 over a period of 1146 days during
the course of its mission lifetime \citep{Perryman97,vanLeeuwen07}. Analysis of these
{\it Hipparcos} data show that they have a 1$\sigma$ RMS scatter of
0.018 mag and a mean measurement uncertainty of 0.012. The data are
thus consistent with HD~6434 being photometrically stable at the
$\sim$1\% level. 

\subsection{Photometry from CTIO}
The TERMS team first monitored HD~6434 with the CTIO 1.0m telescope and Y4KCam CCD detector during 2011 Oct 22-30. The observations were conducted off-transit to establish a baseline and were done using the Johnson-Morgan B-band filter. Relative photometry was performed using the TERMS Photometric Pipeline (see Appendix) with respect to one stable comparison star: CD -40 236. Note we had originally identified two references stars within the science frames, but found that the CD -40 237 was a variable star and therefore unusable. The data have a 1$\sigma$ RMS of 0.007 mag. 

Additional time was required in order to monitor the target during the predicted transit window. Time was allotted on the CTIO 0.9m telescope during 2012 Oct 18-20. The same reference star and filter were used. Due to poor weather, one additional night was needed to determine whether HD~6434 transited during the predicted transit window. The night of 2014 Sep 22 was purchased on the CTIO 0.9m telescope and observations were conducted by David James, again using the same reference and filter. The CTIO 0.9m data have a 1$\sigma$ RMS of 0.016 mag.

\begin{figure}
  \includegraphics[width=9cm]{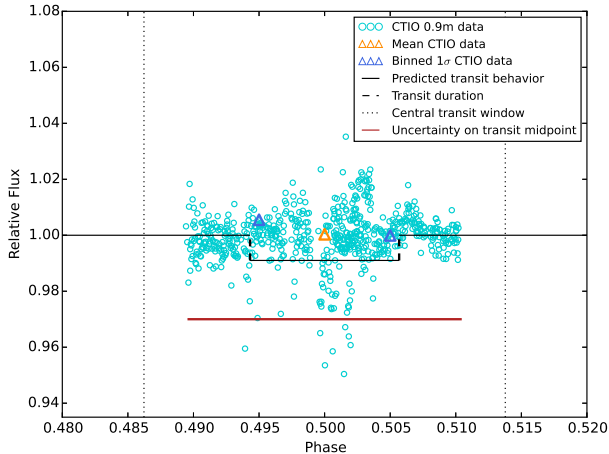}
  \caption{All of the phase folded photometric data taken on the CTIO 0.9 m telescope during the predicted central transit window (dashed lines). The predicted transit behavior is given by the solid black line, demarcating the transit duration with dashed lines, and a predicted transit depth of 0.9\%. The uncertainty on the transit midpoint is given by the solid red line. The mean and binned 1$\sigma$ data are given by the orange and dark blue triangles, respectively. The uncertainty on the transit midpoint and the depth are given by the red solid line. 
}
  \label{phot}
\end{figure}

\subsection{Photometry and Planetary Transit}
\label{transit}

The phase-folded photometric results are given in Fig. \ref{phot}, where all of the data within the window were taken on the CTIO 0.9m telescope. The solid black lines shows the predicted transit behavior of the planet, including the transit duration (0.01 phase or 0.25 days) as dashed black lines (see Table \ref{planet}) at a predicted depth of 0.9\%. These properties were determined using the stellar (Table \ref{starparams}) and orbital planetary ephemerides (Table \ref{planet}). The central transit window (0.031 phase or 0.69 days) was calculated using the error on the stellar radius and improved planetary properties, given by dotted lines. The uncertainty on the transit midpoint (0.021 phase or 0.46 days) is given by the solid red line. 

To verify the trend in the 657 observations within the transit window, we plot the mean (0.9994 mag) as an orange triangle. Additionally, we have divided the data into two 1$\sigma$ bins,
which translates to a width 0.01 in phase-space,
 and determined the means of each of those bins, 1.0055 mag and 0.99994 mag, given as dark blue triangles, respectively.  The means of the two 1$\sigma$ bins, in conjunction with the overall mean of the data, serve to verify the comparatively stable relative flux observations seen within the transit window.
 The data has a 1$\sigma$ RMS scatter of 0.009 mag, which would have captured a transit event with significant certainty.  We can rule out the possibility of a HD~6434b transiting in front of the host star to a depth of 0.9\%. Additionally, given that these measurements cover 75.4\% of the predicted central transit window, centered on the transit window, without a significant decrease in the relative flux, we can rule out a grazing transit to a depth of 1.6\% due to the limitations of the impact parameter.

\section{Conclusions}
Using 137 RV measurements, including 59 new measurements from the {\sc CORALIE} spectrometer, we have improved on the Keplerian orbital solution for HD 6434b in order to determine highly precise transit ephemeris for the planetary system. We used the accuracy offered by the TERMS project, which is unmatched by individual ground-based telescopes, in order to determine if the exoplanet transited in front of the host star. We observed HD6434 using the CTIO 0.9m and 1.0m telescopes, in addition to published {\it Hipparcos} data. The photometric data from CTIO were analyzed using the TERMS Photometry Pipeline, which was developed by our team to ensure accurate and consistent photometric results. The planet had a 4\% transit probability and 0.9\% transit depth, and our photometric observations covered 75.4\% of 0.69 day 1$\sigma$ transit window. The mean relative flux was 0.9994 mag and the average within two 1$\sigma$ bins was 1.0055 mag and 0.99994 mag. Because the data only had a 1$\sigma$ RMS scatter of 0.009 mag, we are able to confidently rule out a transit, a dispositive null result, to a depth of 0.9\% and a grazing transit to a depth of 1.6\%. 

Because of the additional data acquired, we were able to confidently determine the stellar radius (1.14 $\pm$ 0.05 R$_{\odot}$) and mass (0.89 $\pm$ 0.04 M$_{\odot}$). 
Per our discussion in Section \ref{sec:abund}, we note that we have adopted the renormalized SME measurement of [Fe/H] = $-$0.61 $\pm$ 0.07 dex. Despite the large discrepancies in the [Fe/H] determinations from other datasets, we are not signifying that the data acquired here is any better than other datasets, only that the techniques are different \citep{Hinkel14}. We have chosen to use the SME determination so that our results may be consistent with the associated T$_{\text{eff}}$\,\, and $\log$(g)\, measurements, for which there is a known degeneracy \citep{Torres12}. In addition, we found that the [Fe/H] determinations both from here and within the {\it Hypatia Catalog}, see Table \ref{starparams}, are relatively consistent to within error. Therefore, the mass-radius relationship within this paper is consistent with other literature results. Additionally, the larger RV dataset presented here helps to mitigate the notion that the planetary signature is due to stellar rotation modulation \citep{Mayor04}. Meaning, despite the lack of a transit, the TERMS project has allowed for highly precise characterization of bright planetary systems by collaborating with people all over the world. The TERMS team continues to improve upon the characterization and understanding of both stars and their orbiting exoplanets.

\section*{Acknowledgements}
The authors would like to thank Jason Wright for his useful discussions about this planetary system as well as Eric Mamajek for his help classifying the spectral type of the stellar host. N.R.H. would like to thank CHW3 and acknowledge financial support from the Nexus for Exoplanet System Science and the National Science Foundation through grant AST-1109662. T.S.B. acknowledges support provided through NASA grant ADAP12-0172.

\begin{appendix}

\renewcommand{\thefigure}{A\arabic{figure}}

\setcounter{figure}{0}

\section{TERMS Photometry Pipeline}
\label{tpp}
The Terms Photometry Pipeline (TPP) is a photometric data reduction and analysis tool based in IDL and developed by a member of the TERMS team to efficiently process data from numerous telescopes. The graphic user interface (or GUI) allows for both automation and on-the-fly user input while quickly processing the data. Viewing the data in high-resolution (as possible) is done by scaling the contrast of images based on the {\sc Z-Scale} algorithm from IRAF. The header for each science frame can be displayed to examine the information within, and the zoom function allows for in-depth examination in order to identify pathological data points such as bad pixels and cosmic ray hits.

\subsection{TPP Data Reduction}
\begin{figure*}
  \begin{tabular}{cc}
    \includegraphics[width=0.47\textwidth]{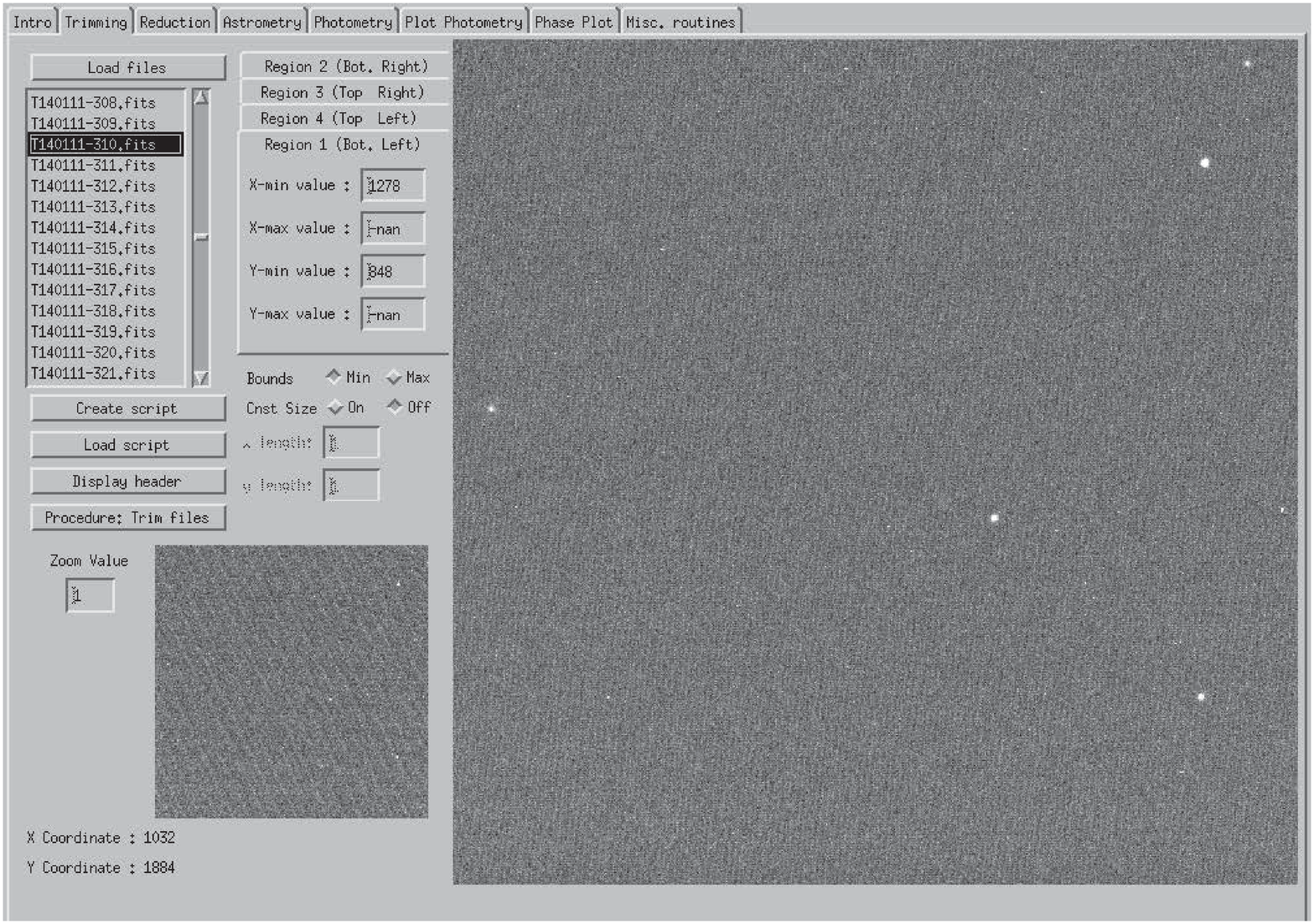} & \includegraphics[width=0.47\textwidth]{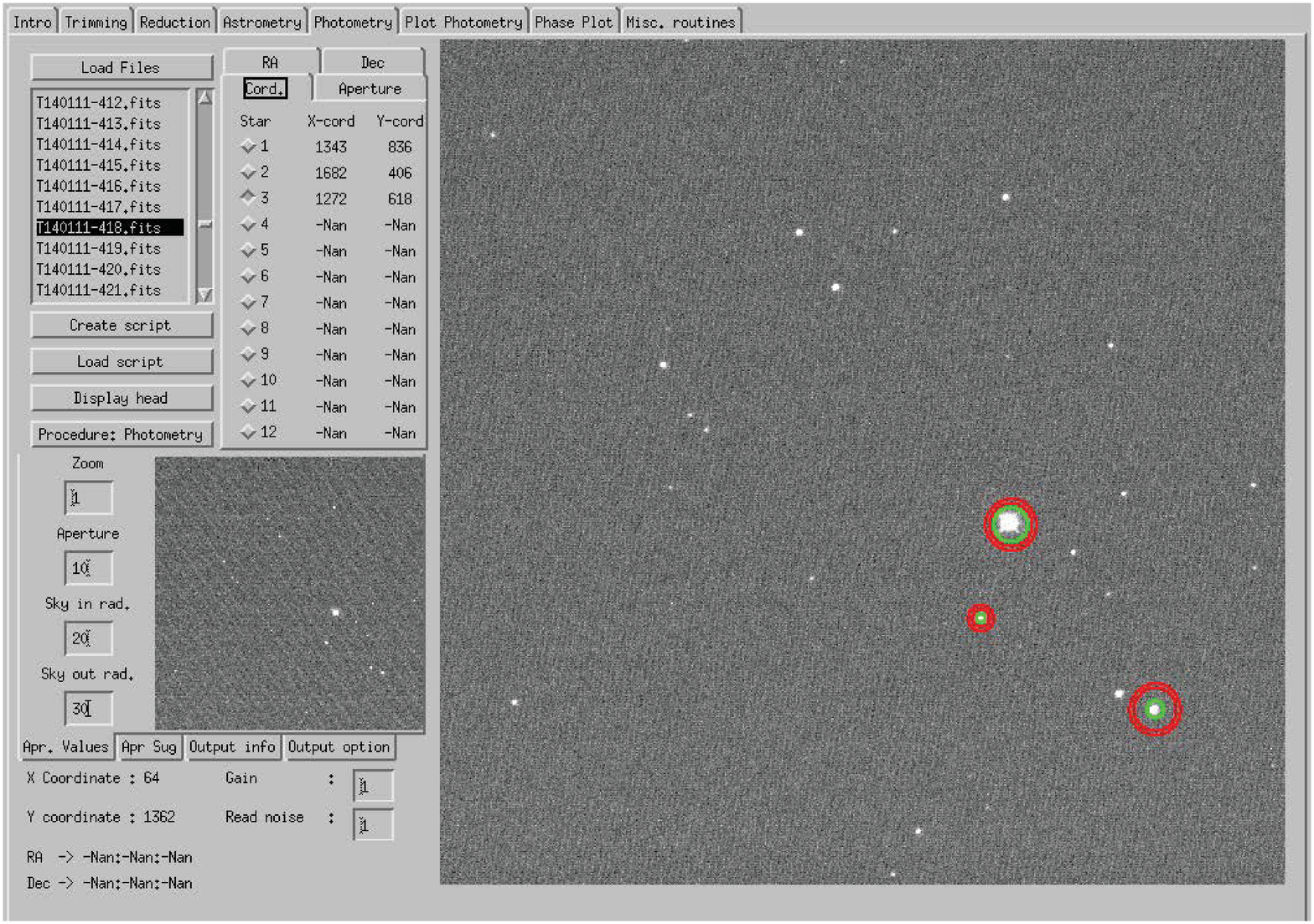} \\
  \end{tabular}
  \caption{{\it Left}: Reduction tab of the GUI. List of the science frames is on the top left panel. A small overview window is available on the bottom left, and a larger window which can be moved/zoomed is on the right. Additional options are scattered through out the GUI {\it Right}: Plot showing the Aperture Suggestion Routine, which randomly selects science images and performs aperture photometry per a curve-of-growth analysis over a range of apertures. The user can systematically and consistently choose appropriate values for the data before final reduction.}
    \vspace{5mm}
 \label{img:terms_trim_phot}
\end{figure*}

The reduction performed by the TPP is done in two steps. Initially, the over-scan region of each science frame is trimmed. If the image is produced from multiple CCD chips, a mosaic can be created from the individual frames. The current chip geometry options are 1x1, 1x2, 2x1 and 2x2. Once the user has specified the bounds for the trimming, the frames are cropped, stitched together, and saved to a sub folder such that the original frames are never changed. An example of the GUI is shown in Fig. \ref{img:terms_trim_phot} (left). Once the reduction preferences are set up, the options can be saved to a text file and loaded in the future.

The second part of the reduction creates the master-bias and -flat frames, then removes their effects from the science images containing the object(s) of interest. The user loads trimmed images\footnote[1]{untrimmed images are allowable, albeit not recommended} and is able to filter the entire data set based on two sets of keywords. For example, by first sorting by the header-keyword ``flat'' will reduce the file list to only frames containing sky/dome flats. A second sort, for example on the keyword ``R'', can then be used to further reduce the list of files that contains dome/sky flats acquired with the R filter. Each frame can be examined individually, added or removed from the list, such that only high quality data are kept for reduction. The user is then able to create a master-bias and -flat frames (or load them if they have already been made). The master-bias is created by taking the average of all the trimmed bias frames selected. After the bias has been subtracted, the trimmed flats are normalized and averaged to produce the master-flat. After creating/loading the master-bias and -flat frames, the user can use the GUI to specify a new file list in order to specify the target frames. The science frames are reduced by subtracting the master-bias and dividing by the master-flat.

\subsection{TPP Photometric Routine}
The photometric routine utilizes the {\sc APER} code, which is based on previously developed work per {\sc DAOPHOT} and its implementation, in IDL. The stars observed with the TERMS project are bright and therefore are often purposefully unfocused in order to avoid having saturated pixels in the object profile, making point-spread function fitting not applicable. The user specifies the location of the target and reference stars by using mouse-clicks, see Fig. \ref{img:terms_trim_phot} (right). The center of the star is found though a centroid algorithm, where pixels containing less then the median value of the total count are excluded from the calculation in order to remove the effects of the background. The calculation for the centroid is analogous to one for center of mass:
\begin{equation}
C_{r}=\frac{1}{F_{tot}} \sum_{i=0}^{n} F_{i}r_{i}
\end{equation}
where $C$ is the center, $F$ is the pixel value for frame $i$, $F_{tot}$ is the total value of all pixels in the region once the background has been removed, and $r$ is the pixel coordinate ($x, y$) on the CCD.  We have determined that such an approach works well for non-crowded fields. Photometry may be performed for up to 12 objects (target $+$ reference stars) across the entire image dataset.

\begin{figure}
\centering
\includegraphics[width=0.55\textwidth]{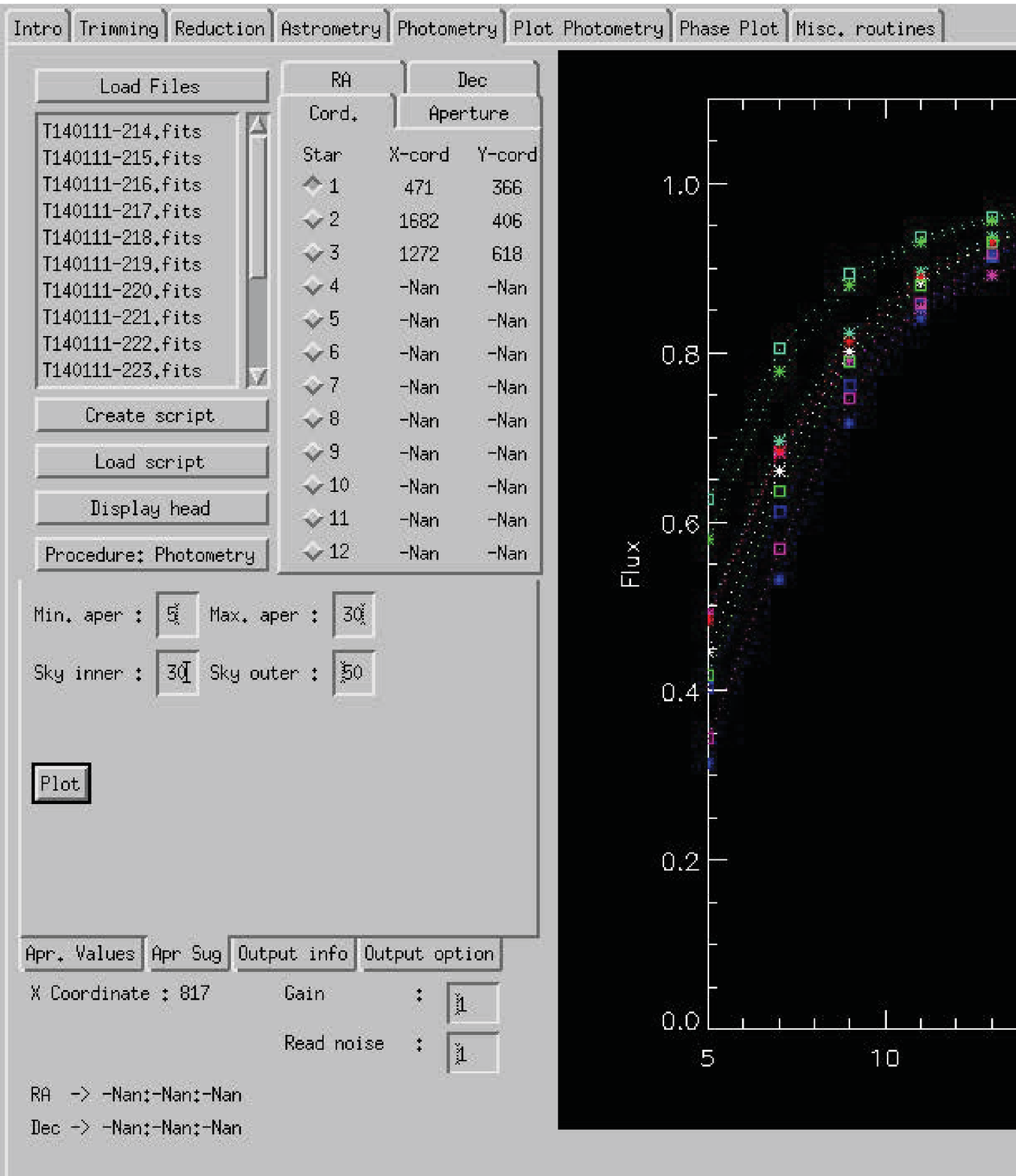}
\caption{Plot showing the Aperture Suggestion Routine, which randomly selects science images and performs aperture photometry per a curve-of-growth analysis over a range of apertures. The user can systematically and consistently choose appropriate values for the data before final reduction.}
\label{fig:terms_aper}
\end{figure}

Aperture photometry, used for subtracting the background signal, is then performed about the center. Ideal aperture and sky annuli are obtained with the Aperture Suggestion Routine built into the TPP. Once initiated with an aperture range, inner sky radius, and outer sky radius, the TPP selects random images and linearly changes the pixel size of the aperture, producing flux as a function of aperture radius. This approach is  resembles to a curve-of-growth analysis (Fig. \ref{fig:terms_aper}), such that when the photon count levels off at a certain point, an increase in aperture size only introduces additional counts from the background and not the actual target. The Aperture Suggestion Routine allows for aperture, inner-, and outer-sky radii to be determined consistently for all science frames.

\subsection{Additional Functions}
The TPP has two built-in routines which are not necessary for the data analysis but can be useful in particular situations. The first routine is a simple IDL wrapper for astrometry.net \citep{Hogg08}. This enables the user to obtain accurate RA and DEC J2000 coordinates, which at times may be unknown, and use them to determine the location of potential target stars.

The secondary supplementary routine available is a plotting function, implemented in order to provide the observer at the telescope with fast analysis of the data as they are being obtained. The plotting routine allows the user to graph parameters like flux, airmass, x/y pixel coordinates, time, as well as any customizable variables, for up to 12 stars. Furthermore, relative photometry can be performed on any target, using any combination of up to 12 stars within the data set.

\subsection{Testing the TPP}
We have thoroughly tested the TPP against published datasets, both from within the TERMS team and from external sources, to verify that the photometric analysis accurate and repeatable. Furthermore, given the Aperture Suggestion Routine, our results are also easily reproducible and do not depend on the person reducing the data. We have used the TPP while ``in the field," reducing data taken on previous nights in order to inform upcoming observations, and have had excellent success. The TTP has been tested and used with telescopes from Lowell Observatory (not shown here) as well as the CTIO 1.0m and 0.9m telescopes. Please contact the author if you would like a copy of the TPP for these or additional telescopes.

\end{appendix}

\nocite{*}
\bibliographystyle{apj}

\end{document}